\documentclass[sigconf]{acmart}

\usepackage{multirow}
\AtBeginDocument{%
  }
\usepackage{bm}

\copyrightyear{2025}
\acmYear{2025}
\setcopyright{acmlicensed}
\acmConference[MM '25]{Proceedings of the 33rd ACM International Conference on Multimedia}{October 27--31, 2025}{Dublin, Ireland}
\acmBooktitle{Proceedings of the 33rd ACM International Conference on Multimedia (MM '25), October 27--31, 2025, Dublin, Ireland}
\acmDOI{10.1145/3746027.3755656}
\acmISBN{979-8-4007-2035-2/2025/10}
\begin{document}



\title{Controllable Video-to-Music Generation with Multiple Time-Varying Conditions}

\author{Junxian Wu}
\affiliation{%
  \institution{Zhejiang University}
  \city{Hangzhou}
  \country{China}}
\email{wujunxian@zju.edu.cn}

\author{Weitao You}
\authornote{Corresponding Author.}
\affiliation{%
  \institution{Zhejiang University}
  \city{Hangzhou}
  \country{China}}
\email{weitao_you@zju.edu.cn}

\author{Heda Zuo}
\affiliation{%
  \institution{Zhejiang University}
  \city{Hangzhou}
  \country{China}}
\email{zuoheda@zju.edu.cn}

\author{Dengming Zhang}
\affiliation{%
  \institution{Zhejiang University}
  \city{Hangzhou}
  \country{China}}
\email{dmz@zju.edu.cn}

\author{Pei Chen}
\affiliation{%
  \institution{Zhejiang University}
  \city{Hangzhou}
  \country{China}}
\email{chenpei@zju.edu.cn}

\author{Lingyun Sun}
\affiliation{%
  \institution{Zhejiang University}
  \city{Hangzhou}
  \country{China}}
\email{sunly@zju.edu.cn}

\renewcommand{\shortauthors}{Junxian Wu, Weitao You, Heda Zuo et al.}

\begin{abstract}
Music enhances video narratives and emotions, driving demand for automatic video-to-music (V2M) generation. However, existing V2M methods relying solely on visual features or supplementary textual inputs generate music in a black-box manner, often failing to meet user expectations. To address this challenge, we propose a novel multi-condition guided V2M generation framework that incorporates multiple time-varying conditions for enhanced control over music generation. Our method uses a two-stage training strategy that enables learning of V2M fundamentals and audiovisual temporal synchronization while meeting users’ needs for multi-condition control. In the first stage, we introduce a fine-grained feature selection module and a progressive temporal alignment attention mechanism to ensure flexible feature alignment. For the second stage, we develop a dynamic conditional fusion module and a control-guided decoder module to integrate multiple conditions and accurately guide the music composition process. Extensive experiments demonstrate that our method outperforms existing V2M pipelines in both subjective and objective evaluations, significantly enhancing control and alignment with user expectations.
\end{abstract}



\maketitle

\section{Introduction}
Music is essential for enhancing the emotional and narrative impact of videos, capturing the audience's attention and interest~\cite{ma2022research, dasovich2022exploring}. Traditional video soundtracks rely on manual synchronization of music with video, a process that is both cumbersome and time-consuming. V2M generation addresses this challenge, aiming to generate music that aligns semantically and temporally with video content. Nevertheless, despite significant progress in V2M, challenges persist in achieving robust controllability and in fully meeting user expectations for emotional and musical alignment~\cite{ji2025comprehensive}.

Firstly, one of the key challenges in V2M generation is that a single video can correspond to multiple suitable music tracks, making it hard to ensure the generated music aligns with user expectations. While most existing studies~\cite{di2021video, zhuo2023video, zhang2025sonique, lin2024vmas} focus on music generation by extracting video features, such as semantic, motion, or color information, they often overlook the specific requirements of individual users, causing the generated music to fall short of their expectations. Some studies~\cite{su2024v2meow, liu2023m} incorporate textual input for additional control. However, text often fails to convey nuanced emotional dynamics within the video, resulting in music that does not accurately reflect its dynamic mood. Furthermore, textual features lack temporal continuity, limiting their ability to represent the dynamic nature of generated music. Even when temporal information is incorporated into the text, it typically requires detailed and complex descriptions, making the process time-consuming.

To enhance controllability of music generation, multi-condition control has been extensively explored in text-to-music (T2M) generation. Most existing methods~\cite{melechovsky2023mustango, wu2024music, copet2024simple} integrate various control factors, including chords, melody, and rhythm, which preserve fundamental music features. However, they often neglect higher-level attributes, such as semantics and emotion. In contrast, models like~\cite{rouard2024audio} utilize style conditioners and abstract features for music generation, but they fail to consider musical elements, and the corresponding latent features lack precise time-varying controls.

\begin{figure}[h]
  \centering
  \includegraphics[width=\linewidth]{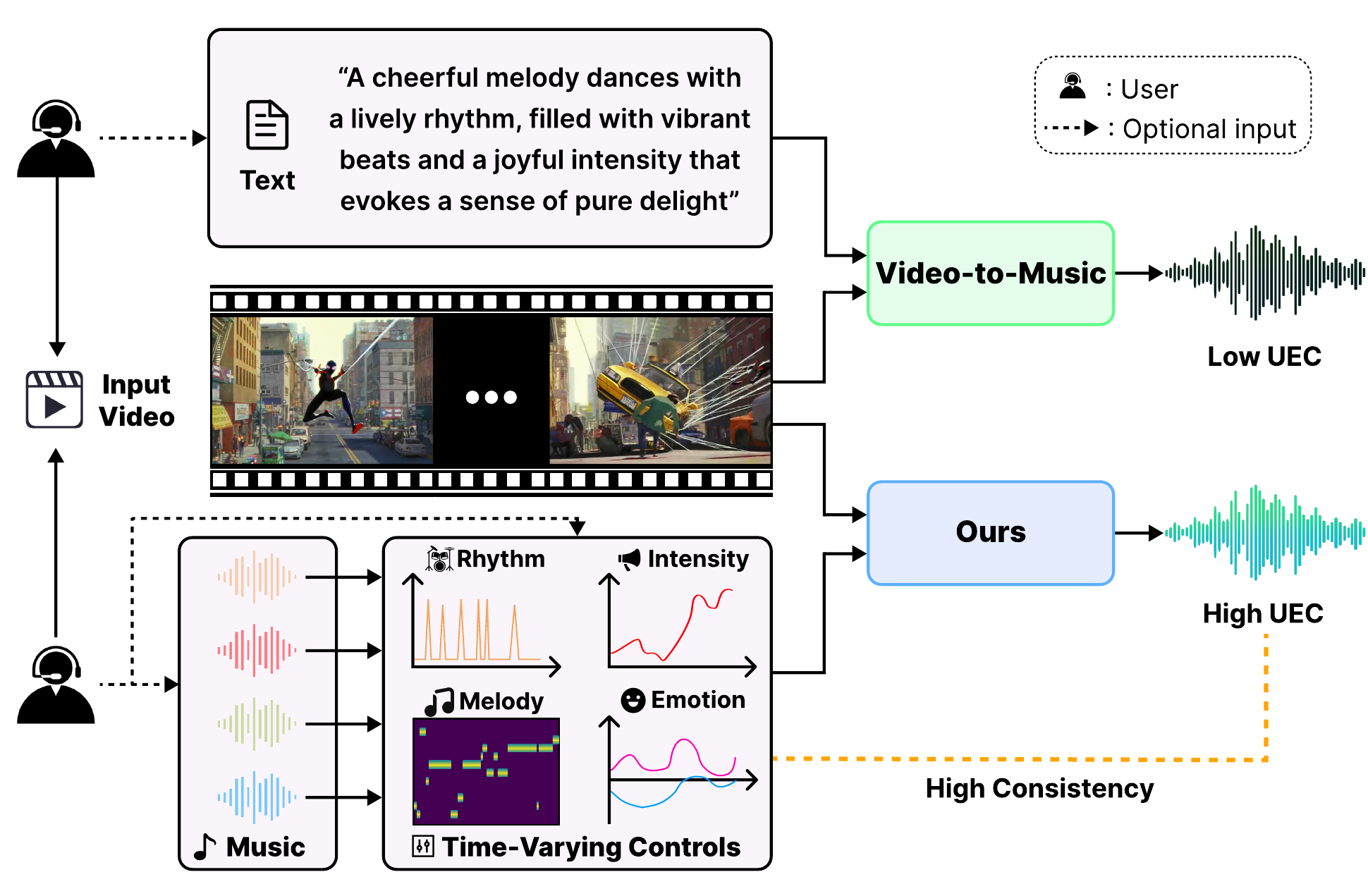}
  \caption{Comparison of V2M generation frameworks between existing models and ours. Unlike models that rely solely on video and optional text, our model uses video and multiple optional time-varying conditions. This results in higher user expectation conformity (UEC), as users can adjust specific dynamic features to suit their preferences.}
  \label{Comparison}
\end{figure}

Moreover, fixed temporal alignment strategies limit the effectiveness and flexibility of V2M generation models, hindering dynamic and controllable music generation. Previous methods like~\cite{zhuo2023video, kang2024video2music} primarily focus on the global features of the entire video clip, while some other models~\cite{su2024v2meow, zuo2025gvmgen} emphasize frame-level video features. Tian et al.~\cite{tian2024vidmuse} combine local and global visual features. However, these fixed temporal alignment strategies fail to dynamically adapt to varying temporal contexts, thereby limiting the model’s flexibility and reducing its precision in generating music that aligns with the diverse and evolving content of the video.

To address the challenges of limited controllability and temporal alignment in V2M generation, and inspired by the T2M domain~\cite{lan2024musicongen}, we propose a novel multi-condition guided V2M generation framework. As shown in Figure~\ref{Comparison}, unlike previous methods that either rely solely on visual features or limited textual control, our method introduces four time-varying conditions: beat, melody, intensity, and emotion. These conditions offer fine-grained control over music generation, enabling users to manipulate specific aspects: beat governs rhythmic structure, melody shapes musical coherence and harmony, intensity modulates energy levels, and emotion influences the expressive quality of the music. By incorporating multiple dynamic conditions, our framework enables more precise and flexible music generation, thereby significantly improving controllability.

Building upon our proposed framework, we introduce a two-stage training strategy and develop the first V2M generation model with multiple time-varying controls. This strategy enables the model to first acquire a foundational understanding of V2M generation and temporal alignment through pretraining, and then integrate multiple conditions into music generation via fine-tuning. Specifically, in the first stage, we employ a video feature aggregation module to determine the overall tone of the music and a progressive temporal alignment mechanism for more flexible feature alignment. To facilitate this process, we propose a fine-grained feature selection module that retains only the most relevant features. In the second stage, we design a dynamic conditional fusion module that assigns feature weights dynamically based on their relevance to the video, followed by a control-guided decoder module that leverages the fused features to guide the music composition. This stage refines the decoder's output by adjusting the generated music based on time-varying conditions, ensuring it remains dynamically aligned with the input and is contextually appropriate for the video.

The main contributions are summarized as follows:
\begin{itemize}
    \item We propose a novel V2M generation framework with multiple time-varying controls, including melody, beat, intensity, and emotion, enabling more precise and user-controllable music generation.
    \item For this framework, we introduce a two-stage training strategy capable of achieving flexible temporal alignment and dynamically integrating multi-condition control.
    \item Extensive experiments show that our method outperforms the current state-of-the-art in both subjective and objective evaluations, achieving significant improvements in controllability and better alignment with user expectations.
\end{itemize}

\begin{figure*}[h]
    \includegraphics[width=\textwidth]{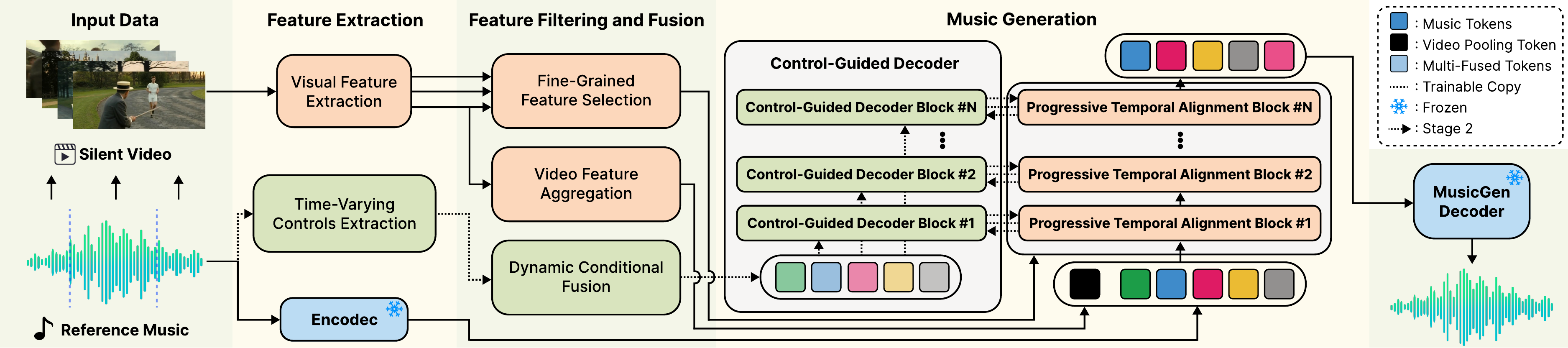}
    \caption{The main architecture of the proposed model, consisting of a two-stage process: pre-training and fine-tuning.}
    \label{fig:1}
\end{figure*}

\section{Related Works}
\subsection{Video-to-Music Generation}
\par Recently V2M generation has garnered significant attention. Early methods~\cite{gan2020foley, su2020multi, zhu2022quantized} focus on generating music from human movements but are inapplicable to more general videos. CMT~\cite{di2021video} first introduced the V2M task by leveraging video motion features to predict music features. Video2Music~\cite{kang2024video2music} and V-MusProd~\cite{zhuo2023video} incorporate various visual features to generate music. However, these methods produce monotonous symbolic music. Models like GVMGen~\cite{zuo2025gvmgen}, VidMuse~\cite{tian2024vidmuse} extract hidden video features and use them for waveform music generation. However, they consider only visual features and fail to address users' specific preferences, leading to music that may not align with user expectations. Although V2Meow~\cite{su2024v2meow} and Diff-BGM~\cite{li2024diff} use visual-text pairs for music generation, text is limited in accurately conveying dynamic temporal information and variations. Moreover, their fixed alignment mechanisms fail to dynamically adapt to varying temporal contexts, including frame-level alignment with either the whole music~\cite{zuo2025gvmgen} or music frames~\cite{li2024muvi}, alignment of combined local and global features with the entire music sequence~\cite{tian2024vidmuse}, and segment-aware feature alignment~\cite{li2024diff}. Therefore, we incorporate multiple time-varying conditions for fine-grained control, and propose a flexible temporal alignment mechanism for precise feature synchronization.

\subsection{Music Generation with Multiple Conditions}
For more controllable music generation, various control signals are proposed, such as images~\cite{wang2023continuous, liang2024drawlody}, videos~\cite{di2021video, zuo2025gvmgen}, audio~\cite{choi2023pop2piano}, natural languages~\cite{agostinelli2023musiclm, evans2024fast} or their combinations~\cite{chowdhury2024melfusion, liu2023m}. However, control over multiple musical elements, which are crucial for precise music generation, has only been extensively explored in T2M tasks. Mustango~\cite{melechovsky2023mustango} uses a diffusion model to guide music generation towards input tempo, key, chords, and general textual description. Music Controlnet~\cite{wu2024music} employs a diffusion model architecture and the adapter-based conditioning mechanism of ControlNet~\cite{zhang2023adding} to manipulate text, melody, dynamics, and rhythm conditions. Musicongen~\cite{lan2024musicongen} presents a Transformer-based T2M generation model that follows rhythm and chord conditions. These methods utilize musical features but overlook higher-level attributes, limiting the richness of musical expression. In contrast, Rouard et al.~\cite{rouard2024audio} use style conditioners and abstract features to generate music, but they do not consider musical elements, and the abstract features lack precise temporal control. Therefore, we introduce time-varying musical and emotional elements in V2M generation, and propose a dynamic conditional fusion module and a control-guided decoder module to better integrate multiple conditions into the framework.

\section{Methods}
\subsection{Problem Formulation}
In V2M generation with multi-condition control, our goal is to learn a conditional generative model $p(M\,|\,V_{\text{Ref}}, C)$ over generated music $M$, given a reference video $V_{\text{Ref}}$ and a set of time-varying controls $C$ (i.e., rhythm $C_{\text{Rhy}}$, intensity $C_{\text{Int}}$, melody $C_{\text{Mel}}$, and emotion $C_{\text{Emo}}$). The reference video $V_{\text{Ref}} \in \mathbb{R}^{t \times f_v \times C_v \times H \times W}$ serves as the main input, where $t$, $f_v$, $C_v$, $H$, and $W$ represent the duration, video frame rate, number of channels, frame height, and frame width, respectively. The conditions $C$ are denoted as $\mathbb{R}^{t \times f_m \times D}$, with $f_m$ and $D$ being music sample rate and feature dimension of the corresponding control. Music is represented as quantized codes $M\in \mathbb{R}^{t \times f_m \times K}$, which are derived from Encodec~\cite{defossez2022high} with $K$ codebooks.

\begin{figure*}[h]
    \includegraphics[width=\textwidth]{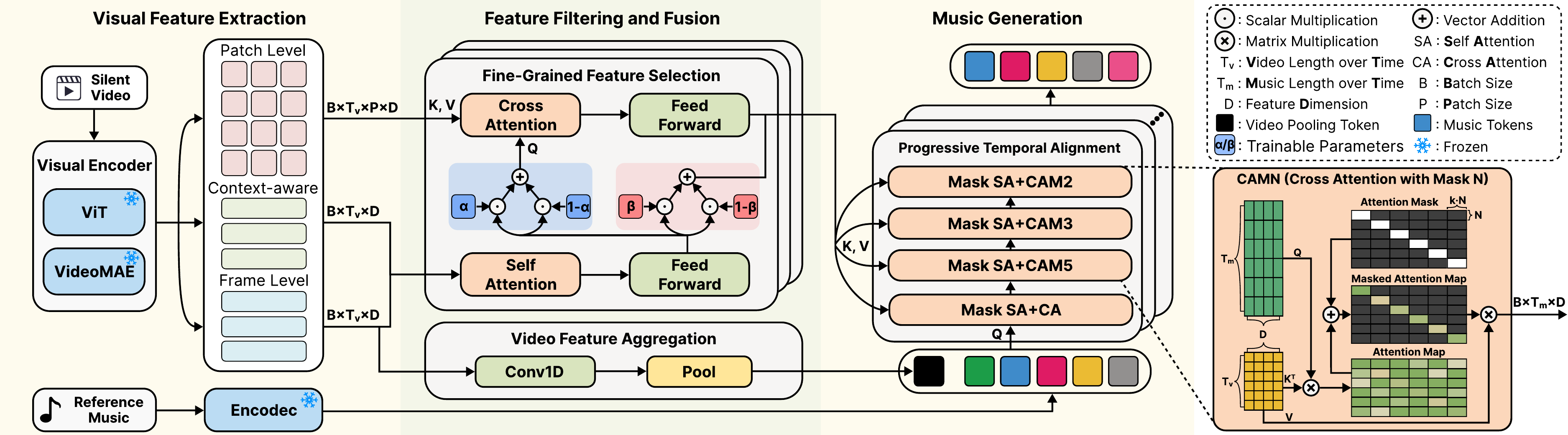}
    \caption{Video-to-music generation pre-training stage.}
    \label{fig:2}
\end{figure*}

\subsection{Method Overview}
\par The main architecture of the proposed method is shown in Figure~\ref{fig:1}. Our method involves two training stages: V2M generation pre-training and multi-condition control fine-tuning. 

\par In the first stage, a hierarchical visual feature extractor processes the input video to derive three complementary features: patch-level fine-grained image features, frame-level visual features, and context-aware visual features. The Video Feature Aggregation (VFA) module utilizes the frame-level features to form video-level semantic features that serve as the foundation for the overall musical tone. Simultaneously, a Fine-Grained Feature Selection (FGFS) module leverages frame-level and context-aware visual features to filter raw patch-level fine-grained details, yielding a refined representation of music-related features. Finally, a Progressive Temporal Alignment Attention (PTAA) module is introduced to enhance flexibility of music generation using the refined features. Consequently, our model can generate music that corresponds to the diverse types and rhythmic variations present in the input videos.

\par In the fine-tuning stage, time-varying controls and their associated modules are introduced to refine the music composition. The Dynamic Conditional Fusion (DCF) module assigns time-varying feature weights to integrate multiple conditions, ensuring effective multi-condition guidance. Subsequently, the Control-Guided Decoder (CGD) module refines the generated music by adjusting the decoder’s output based on the fused conditions. By incorporating these two modules, the model composes video-conditioned music to align with the visual input while better integrating user-specified, fine-grained conditions. We detail each module below.

\subsection{Video-to-Music Generation Pre-training}
\par As shown in Figure~\ref{fig:2}, the pre-training consists of three parts: 1) aggregating frame-level features into a unified video representation to guide the musical theme; 2) selectively filtering fine-grained visual features to facilitate subsequent temporal alignment; and 3) progressively aligning music with the video content. 

\par Before introducing the modules, we first discuss the visual feature extraction, which underpins their functionality. We extract three distinct video features: patch-level fine-grained image features, frame-level visual features via CLIP~\cite{radford2021learning}, and context-aware visual features via VideoMAE V2~\cite{wang2023videomae}. These features capture varying aspects of the video, from spatial details to broader contextual relationships, providing a comprehensive video representation.

\par \textbf{VFA module}. To guide music generation with a comprehensive understanding of the video, we propose the VFA module, which extracts frame-level features and aggregates them into a unified visual representation. This representation serves as the foundation for determining the overall musical theme, providing essential, compact, yet expressive information at the start of the generation process. Specifically, we utilize frame-level visual features $V_{f}\in \mathbb{R}^{T \times D}$ to capture both local and global dependencies within the visual data. Subsequently, a one-dimensional convolutional layer (Conv1D), followed by a temporal pooling layer, aggregates these features into a compact feature vector $V_{\text{agg}}\in \mathbb{R}^{1 \times D}$ that encapsulates the video's overall features. This vector is then positioned before the first music token, ensuring that the subsequent generation process is conditioned on a meaningful and holistic video representation.

\par \textbf{FGFS module}. This module is designed to facilitate fine-grained temporal alignment by retaining only video features relevant to music generation. Raw video features often contain redundant information that hinders alignment with the generated music. To address this, we filter and refine fine-grained visual features by utilizing both local features and global structure, ensuring adaptation to the varying rhythms of different videos and coherence in music.

\par Specifically, we first integrate frame-level and context-aware visual features via self-attention~\cite{vaswani2017attention} to capture both local and global dependencies within the video. Inspired by~\cite{chowdhury2024melfusion}, we introduce learnable parameters $\alpha$ and $\beta$, which modulate feature contributions before further fusion. In the fusion process, we use a cross-attention mechanism, where patch-level fine-grained image features, downsampled via a 2D convolutional layer (Conv2D), serve as keys ($\bm{K}$) and values ($\bm{V}$), while the $\alpha$-weighted fusion of frame-level and context-aware features acts as the queries ($\bm{Q}$). The resulting refined features are then concatenated with the $\beta$-weighted fusion output, forming a structured representation for the next block. This hierarchical fusion adaptively preserves fine-grained details while integrating broader contextual information. The $\alpha$-weighted and $\beta$-weighted fusion outputs can be expressed as: 
\begin{equation}
  \bm{K}_{i}^{C} = \alpha_{i}\bm{K}_{i}^{C}+(1-\alpha_{i})\bm{K}_{i}^{F}, ~
  \bm{V}_{i}^{C} = \alpha_{i}\bm{V}_{i}^{C}+(1-\alpha_{i})\bm{V}_{i}^{F}
\end{equation}
\begin{equation}
  \bm{K}_{i}^{F} = \beta_{i}\bm{K}_{i}^{C}+(1-\beta_{i})\bm{K}_{i}^{F}, ~
  \bm{V}_{i}^{F} = \beta_{i}\bm{V}_{i}^{C}+(1-\beta_{i})\bm{V}_{i}^{F}
\end{equation}
where $\bm{K}_{i}^{C}$ and $\bm{V}_{i}^{C}$ are context-aware visual features for the layer $i$ of the module and $\bm{K}_{i}^{F}$ and $\bm{V}_{i}^{F}$ are frame-level visual features. 

\begin{figure*}[h]
    \includegraphics[width=\textwidth]{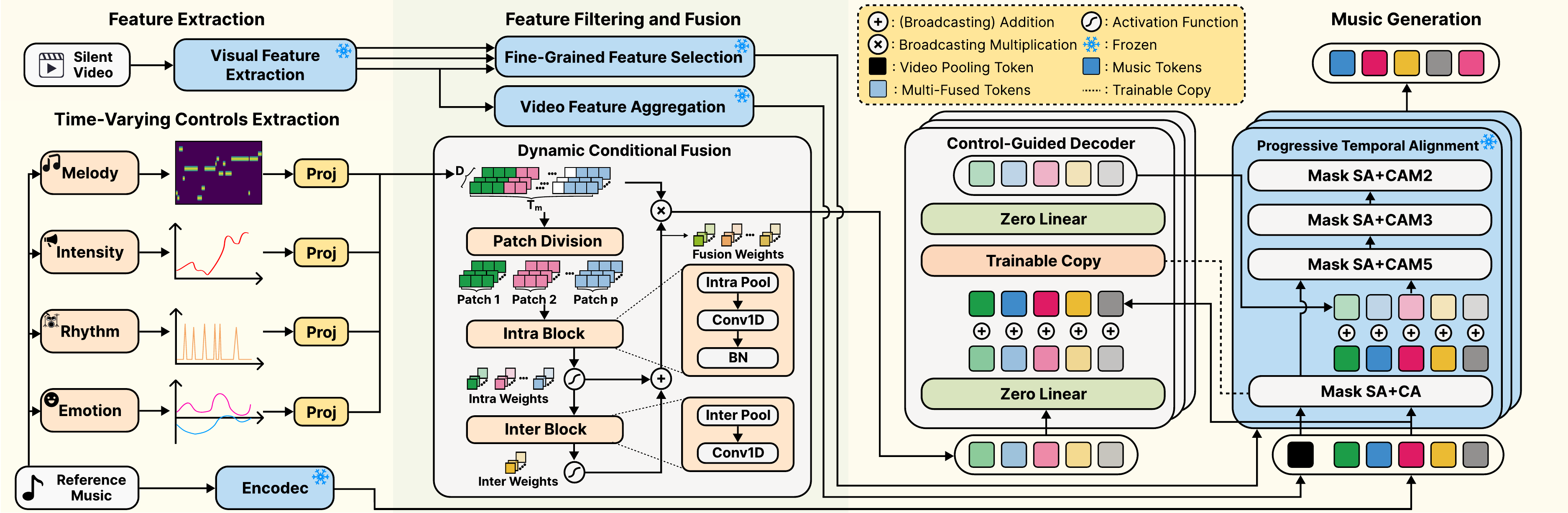}
    \caption{Multi-condition control fine-tuning stage. $\text{T}_\text{m}$ and D represent music length over time and music feature dimension.}
    \label{fig:3}
\end{figure*}

\par \textbf{PTAA module}. To ensure temporal consistency between evolving visual content (e.g., shot transitions, motion dynamics) and music, we propose the PTAA module, which adaptively aligns audio-visual features across multiple temporal resolutions and attention windows. Unlike conventional Transformer-based decoders~\cite{tian2024vidmuse, zuo2025gvmgen} and fixed alignment methods~\cite{li2024diff}, PTAA refines alignment via a hierarchical attention mechanism, which enables the model to capture nuanced local variations while maintaining long-range temporal dependencies. By eliminating fixed alignment priors, it autonomously learns context-sensitive correlations between video rhythm and musical structure, ensuring that generated music not only reflects the video content but also adapts to diverse temporal dynamics.

\par As shown in Figure~\ref{fig:2}, PTAA adopts a decoder-only Transformer architecture~\cite{copet2024simple}. The inputs include fine-grained video features $Y \in \mathbb{R}^{B \times T_{v} \times d}$ and the real music embeddings $X \in \mathbb{R}^{B \times T_{m} \times d}$, where $B$, $T_v$, $T_m$, and $d$ denote the batch size, video frame count, length of the music token sequence, and music feature dimension, respectively. The module generates a music output $X' \in \mathbb{R}^{B \times T_{m} \times d}$ that is flexibly synchronized with the video content. The decoder comprises multiple stacked 4D-Blocks, each consisting of four cascaded masked self-attention and cross-attention layers. Let $X_{0}^{(l)}$ denote the input to the $l$-th 4D-Block. Within each block, the latent representation is iteratively updated through four sequential layers. For layer $j\in\{1, 2, 3, 4\}$, the update equation can be formulated as:
\begin{equation}
    \begin{split}
        X_{j-1}^{(l)} &= \text{LayerNorm}(X_{j-1}^{(l)}+\text{MaskSA}(X_{j-1}^{(l)}, Y, M^{\text{sa}}_{j})), \\
        X_{j}^{(l)} &= \text{LayerNorm}(X_{j-1}^{(l)}+\text{MaskCA}(X_{j-1}^{(l)}, Y, M_{j})),
    \end{split}
\end{equation}
where $\text{MaskSA}$ is the original masked self-attention~\cite{vaswani2017attention}, $M^{\text{sa}}_{j}$ is its corresponding mask, and $\text{MaskCA}$ denotes the masked cross-attention mechanism. The $\text{MaskCA}$ operation employs a scaled dot-product formulation with adaptive temporal masking:
\begin{equation}
\text{MaskCA}(Q_j, K, V, M_j) = \text{softmax}(\frac{Q_jK^T}{\sqrt{d_k}}+M_j)V
\end{equation}
where $Q_j=X_{j-1}^{l} W_q$, $K=Y W_k$, and $V=Y W_v$ are derived from learnable matrices $W_q$, $W_k$, and $W_v$. The mask $M_j\in \mathbb{R}^{T_m \times T_v}$ controls the temporal receptive field by restricting cross-modal interactions to local video segments. Specifically, for each music token at position $i$, only features of $k_j$ adjacent video frames are accessible:
\begin{equation}
    M_j=
    \begin{cases}
        0, & \text{if} \; k_j \cdot \gamma \leq i <  k_j \cdot (\gamma +1) \\
        -\infty, & \text{otherwise},
    \end{cases}
\end{equation}
where $\gamma \in \{0, 1, ...,T/k_j\}$ indexes contextual segments. To enable hierarchical alignment, $k_j$ decreases across layers: the initial layer uses global context ($k_j=\infty$), followed by constrained windows of 5, 3, and 2 seconds in subsequent layers. This multi-scale strategy first establishes coarse-grained structural correspondence and then refines local details, effectively adapting to diverse video dynamics.

\subsection{Musical and Emotional Control Signals}
\par Before the fine-tuning stage, we first introduce four time-varying control signals: rhythm, intensity, melody, and emotion. Inspired by~\cite{wu2024music}, we define two methods for obtaining control signals: extracted controls and created controls. Extracted controls are derived from input audio using feature extraction models without human annotation, represented as $\bm{C} \in \mathbb{R}^{T_m\times D}$, where $D$ denotes the feature dimension of the corresponding signal. In contrast, created controls are curves directly annotated by a music creator, offering greater flexibility and control over the music generation process. Our method is trained with extracted controls, while inference can be conducted using either extracted or created controls. The following sections describe how these controls are obtained.

\par \textbf{Rhythm} ($C_{\text{Rhy}}\in \mathbb{R}^{T_m\times 1}$). For rhythm control, we extract beats and downbeats using a Recurrent Neural Network (RNN)-based beat detector from the Madmom library~\cite{bock2016madmom, bock2016joint}. The detector's outputs are encoded as one-hot embeddings, and then a soft kernel is applied to the downbeats. By summing the beat and downbeat embeddings, we generate the time-varying rhythm control signal $C_{\text{Rhy}}$.

\par \textbf{Intensity} ($C_{\text{Int}}\in \mathbb{R}^{T_m\times 1}$). For intensity control, we compute frame-wise energy from a linear spectrogram using the Librosa library and convert it to decibels. To reduce rapid fluctuations caused by note or percussion onsets, we use a Savitzky-Golay filter~\cite{virtanen2020scipy} with a one-second context window to smooth the signal $C_{\text{Int}}$.

\par \textbf{Melody} ($C_{\text{Mel}}\in \mathbb{R}^{T_m\times 12}$). For melody control, we compute a linear spectrogram and rearrange the energy across frequency bins into 12 pitch classes. To obtain a clearer representation of the melody, we apply an argmax operation to select the most prominent pitch class at each frame, which represents the dominant tone. This results in the frame-wise one-hot encoding $C_{\text{Mel}}$.

\par \textbf{Emotion} ($C_{\text{Emo}}\in \mathbb{R}^{T_m\times 2}$). We represent the emotion control signal using two dimensions: valence (V) and arousal (A)~\cite{russell1980circumplex}. We use a dynamic music emotion recognition model~\cite{zhang2025personalized} to extract the dynamic V and A values $C_{\text{Emo}}$ of the music, enabling us to effectively quantify the emotional features for music generation.

\subsection{Multi-Condition Control Fine-tuning}
\par As shown in Figure~\ref{fig:3}, the fine-tuning framework comprises two components: 1) assigning dynamic weights to integrate multiple time-varying features, and 2) incorporating the fused features into the music decoder. In this process, we freeze all pretrained parameters to prevent catastrophic forgetting. To facilitate subsequent music control, we map the four aforementioned conditions to hidden feature spaces $C_{i} \in \mathbb{R}^{T_m \times D}$, respectively, using linear projection, where $i \in \{1,2,3,4\}$ represents the condition index, $D=D_m/N$, $D_m$ denotes the music embedding dimension, and $N=4$. This projection preserves condition-specific features while ensuring dimensional compatibility for downstream fusion operations.

\par \textbf{DCF module}. This module addresses the limitations of direct concatenation, which cannot adaptively capture the time-varying dominance of specific conditions at distinct timesteps and dynamics of condition importance during generation. To address this issue, DCFM learns adaptive weights across conditions and timesteps to enable effective fusion. We first concatenate all conditions along the feature dimension to form a unified input tensor $C_{\text{in}} \in \mathbb{R}^{T_m \times D_m}$. Then, $C_{\text{in}}$ is partitioned into $N$ non-overlapping patches of fixed length $P$ (with $N=T_m /P$), yielding $C'_{\text{in}} \in \mathbb{R}^{N \times P\times D_m}$.

\par Inspired by the dynamic convolutional network of TVNet~\cite{li2025tvnet}, we propose a patch-aware adaptive fusion framework via temporal-conditioned operations. The output, $C_{\text{out}} \in \mathbb{R}^{T_m \times D_m}$, compatible with the music decoder's input specifications, is formulated as: 
\begin{equation}
    C_{\text{out}}=\alpha \odot C'_{\text{in}}
\end{equation}
where $\alpha \in \mathbb{R}^{T_m \times D_m}$ is a condition-aware weight, and $\odot$ represents element-wise multiplication. To model hierarchical temporal dependencies across patches, we design a dual-context weight generator $\mathcal{G}$ that fully considers both inter-patch and intra-patch interactions: 
\begin{equation}
    \alpha=\mathcal{G}(C'_{in})=\mathcal{F}(x_{\text{intra}})+\mathcal{F}(x_{\text{inter}})
\end{equation}
where $\mathcal{F}(x_{\text{intra}})$ and $\mathcal{F}(x_{\text{inter}})$ handle intra-patch and inter-patch feature fusion, respectively.

\par \textbf{Intra-Patch Feature Fusion}. For the intra-patch block, we use 2D Adaptive Average Pooling on $C'_{in}$ to obtain intra-patch feature embeddings $x_{\text{intra}} \in \mathbb{R}^{D_m \times N}$ and then apply a single-layer Conv1D, denoted as $\mathcal{F}(x_{\text{intra}})$, to $x_{\text{intra}}$. The operations are defined as:
\begin{equation}
\begin{split}
    x_{\text{intra}} &= \text{AdaptiveAvgPool2d}(C'_{\text{in}}), \\
    \mathcal{F}_{\text{intra}}(x_{\text{intra}}) &= \delta(\text{BN}(\text{Conv1D}^{C\rightarrow C}(x_{\text{intra}})))
\end{split}
\end{equation}
Here, $\delta$ and $\text{BN}$ denote activation function and Batch Normalization.

\par \textbf{Inter-Patch Feature Fusion}. For the inter-patch block, we perform 1D Adaptive Average Pooling on $x_{\text{intra}}$ to obtain inter-patch feature embeddings $x_{\text{inter}} \in D_m \times 1$, which aggregate essential features of all patches. Subsequently, we use a single-layer Conv1D, denoted as $\mathcal{F}(x_{\text{inter}})$, on $x_{\text{inter}}$. This process is expressed as:
\begin{equation}
\begin{split}
    x_{\text{inter}} &= \text{AdaptiveAvgPool1d}(x_{\text{intra}}), \\
    \mathcal{F}_{\text{inter}}(x_{\text{inter}}) &= \delta(\text{Conv1D}^{C\rightarrow C}(x_{\text{inter}}))
\end{split}
\end{equation}

\par \textbf{CGD module}. This module is a key component in fine-tuning, integrating multiple conditions into music rearrangement. We adopt the in-attention mechanism from MuseMorphose~\cite{wu2023musemorphose}, applying it similarly to MusiConGen~\cite{lan2024musicongen} and VidMusician~\cite{li2024vidmusician} by incorporating it into the first attention layer of each four-layer transformer block, as shown in Figure~\ref{fig:3}. However, unlike prior approaches that use only linear layers to augment music features by fine-tuning self-attention layers~\cite{lan2024musicongen} or freezing all parameters of the generative backbone~\cite{li2024vidmusician}, we integrate a ControlNet-like module into the in-attention mechanism while keeping the pretrained V2M generation backbone frozen. Inspired by ControlNet's approach of freezing backbone parameters while adding trainable conditional branches to control pretrained text-to-image diffusion models~\cite{zhang2023adding}, we design a parallel control pathway that injects multi-condition fused features into the decoder’s attention hierarchy.

\par Specifically, we create a trainable clone of the first layer of the 4D-Block with parameters $\Theta_c$. This clone accepts an external condition $C\in \mathbb{R}^{T_m\times D}$ (i.e., the output from the DCF module) and connects to the frozen 4D-Block via a zero-initialized D-to-D linear layer $L(\cdot;\cdot)$. In our module, we use two such zero-initialized linear layers with parameters $\Theta_{l1}$ and $\Theta_{l2}$. The complete module computes:
\begin{equation}
M_c=\mathcal{F}(x;\Theta)+L(\mathcal{F}(x+L(C;\Theta_{l1});\Theta_c);\Theta_{l2})
\end{equation}
where $M_c$ is the output of the CGD module, $x$ is the input to the first layer of the 4D-Block, and $\mathcal{F}(\cdot;\Theta)$ represents a trained neutral block, which is the first unit of 4D-Block with parameters $\Theta$.

\par \textbf{Masking Partial Music Controls}. To enable flexible selection of arbitrary combinations or masked conditions among the N control signals, Uni-ControlNet~\cite{zhao2023uni} employs a CFG-like training strategy that randomly drops each control signal $c^{(n)}$ during training. Music ControlNet~\cite{wu2024music} extends this approach by randomly omitting an intermediate segment of the input, forcing the model to restore missing musical cues. However, these methods fail to handle user-specified intermediate segments, which is a common scenario in practice. To address this limitation, we further propose a complementary strategy to enhance robustness and controllability. Specifically, let $\mathcal{I}=\{1,...,N\}$ be the control signal indices, and select a subset $\mathcal{I}'\subseteq\mathcal{I}$ to drop. For retained signals in $\mathcal{I} \backslash \mathcal{I}'$, we randomly sample a pair $(t_{n,a},t_{n,b})\in \{1,2,...,T_m\}^2$ for each of the activate signals with $t_{n,a} < t_{n,b}$, and define the mask as:
\begin{equation}
\begin{split}
    c^{(np)}_t&=
    \begin{cases}
        0, & \text{if} \; t \in \left[t_{n,a},  t_{n,b}\right] \\
        c^{(n)}_t, & \text{otherwise},
    \end{cases} \\
    c^{(nq)}_t&=
    \begin{cases}
        c^{(n)}_t, & \text{if} \; t \in \left[t_{n,a},  t_{n,b}\right] \\
        0, & \text{otherwise},
    \end{cases}
\end{split}
\end{equation}
And the output $c^{(n)}_{out}$ can be expressed as:
\begin{equation}
    c^{(n)}_{out}=
    \begin{cases}
        c^{(np)}, & \text{with} \; \text{probability}  \; p \\
        c^{(nq)}, & \text{with} \; \text{probability}  \; q \\
        c^{(n)}, & \text{with} \; \text{probability}  \; 1-p-q,
    \end{cases}
    \quad \forall n \in \mathcal{I} \backslash \mathcal{I}'
\end{equation}

\section{Experiments}
\subsection{Datasets}
We use the dataset from GVMGen~\cite{zuo2025gvmgen} as our training set, which is a large-scale, high-quality dataset specifically for V2M generation. For evaluation, we randomly sample and combine portions from V2M-bench~\cite{tian2024vidmuse}, SymMV~\cite{zhuo2023video}, and GVMGen test set to construct a comprehensive evaluation set. The vocals of all aforementioned music tracks are removed using a music source separation tool~\cite{rouard2023hybrid}.

\subsection{Implementation Details}
Our implementation involves two training stages: pre-training and fine-tuning. The condition dropout rate is set to 0.5, with $p = 0.05$ and $q = 0.05$. We employ learning rates of 1e-4 and 1e-5 for the first and second stages, respectively. The AdamW optimizer is used with $\beta_1=0.9$, $\beta_2=0.95$, a batch size of 6, and a weight decay of 0.1. A warm-up learning rate is applied to all training stages during the initial 4000 steps. The training lasts for 200 and 50 epochs for the first and second stages respectively on a single NVIDIA A100 card. 

\subsection{Objective Evaluation Metrics}
\par In the traditional V2M generation task, we compute Kullback Leibler Divergence (KLD), Fréchet Audio Distance (FAD)~\cite{kilgour2018fr, gui2024adapting} and Fréchet Distance (FD)~\cite{cramer2019look} to evaluate music fidelity and quality by quantifying the difference between generated and reference audio. CLAP Score~\cite{wu2023large, li2024vidmusician} measures the average cosine similarity between real and generated music features. Density~\cite{naeem2020reliable} assesses the closeness of generated samples to real ones by rewarding those situated in regions densely populated with real samples. For music richness evaluation, we use Diversity~\cite{li2024diff} and Coverage~\cite{naeem2020reliable}. Moreover, we compute ImageBind Score (IB)~\cite{girdhar2023imagebind}, Cross-Modal Relevance (CMR) and Temporal Alignment (TA)~\cite{zuo2025gvmgen} to evaluate music-video correspondence (MV-corr) in both global and temporal aspects.

\par For the multi-condition guided framework, we further assess condition controllability alongside music fidelity and music-video correspondence. For emotion and intensity evaluation, we utilize Pearson Correlation Coefficient (PCC) and Concordance Correlation Coefficient (CCC) to quantify the relationship between the frame-level values of the input and those derived from the generated output, following~\cite{zhang2025personalized, wu2024music}. PCC evaluates linear correlation, while CCC captures both correlation and agreement. For melody evaluation, we use Melody Accuracy (Acc) to assess if the individual pitch labels assigned to each frame are consistent between the supplied melody control and the extracted melody from the output~\cite{wu2024music}. For rhythm evaluation, we adopt Rhythm F1~\cite{davies2009evaluation, raffel2014mir_eval}, which assesses the alignment between beat timestamps derived from the input rhythm control and those from the generated music. More details are provided in the Supplementary Material.

\begin{table*}
    \caption{Objective evaluation of V2M generation.}
    \label{obj:table1}
    \begin{tabular}{ccccccccccc}
        \toprule
         \multirow{2}{*}{Model}&\multicolumn{5}{c}{Music Fidelity}&\multicolumn{2}{c}{Music Richness}&\multicolumn{3}{c}{MV-corr}\\
         \cmidrule(lr){2-6} \cmidrule(lr){7-8} \cmidrule(lr){9-11}
         \multicolumn{1}{c}{}&KLD$\downarrow$&FAD$\downarrow$&FD$\downarrow$&CLAP Score$\uparrow$& Density$\uparrow$& Diversity$\uparrow$& Coverage$\uparrow$ &IB$\uparrow$&CMR$\uparrow$&TA$\uparrow$\\
         \hline
    
         CMT&1.63&8.08&355.46&0.55&0.45&30.67&0.30&0.08&0.64&0.76\\
         M$^2$UGen&1.74&5.41&4.68&0.57&0.49&58.44&0.57&0.13&0.56&0.74\\
         Diff-BGM&1.70&21.74&355.69&0.53&0.12&64.49&0.13&0.06&0.60&0.72\\
         GVMGen&1.01&2.78&2.89&\textbf{0.70}&\textbf{0.89}&64.49&0.73&0.15&\textbf{0.65}&0.69\\
         VidMuse&1.18&4.81&3.73&0.65&0.78&54.59&0.78&0.18&0.61&0.62\\

         Ours&\textbf{0.84}&\textbf{2.19}&\textbf{2.69}&\textbf{0.70}&0.88&\textbf{68.45}&\textbf{0.80}&\textbf{0.19}&0.64&\textbf{0.82}\\
        \bottomrule
    \end{tabular}
\end{table*}

\begin{table*}
    \caption{Objective evaluation of V2M generation with time-varying multi-condition controls, where $^\dagger$ indicates models trained with the two-stage strategy and multi-condition modules, and $^\ddagger$ denotes the variant without the two-stage training strategy.}
    \label{obj:table2}
    \begin{tabular}{cccccccccccccc}
        \toprule
         \multirow{2}{*}{Model}&\multicolumn{2}{c}{Emotion Valence}&\multicolumn{2}{c}{Emotion Arousal}&\multicolumn{1}{c}{Melody}&\multicolumn{2}{c}{Intensity}&\multicolumn{1}{c}{Rhythm}&\multicolumn{2}{c}{Music Fidelity}&\multicolumn{3}{c}{MV-corr}\\
         \cmidrule(lr){2-3} \cmidrule(lr){4-5} \cmidrule(lr){6-6} \cmidrule(lr){7-8} \cmidrule(lr){9-9} \cmidrule(lr){10-11} \cmidrule(lr){12-14}
         \multicolumn{1}{c}{}&PCC$\uparrow$&CCC$\uparrow$&PCC$\uparrow$&CCC$\uparrow$&Acc (\%)$\uparrow$&PCC$\uparrow$&CCC$\uparrow$&F1 (\%)$\uparrow$&KLD$\downarrow$&FAD$\downarrow$&IB$\uparrow$&CMR$\uparrow$&TA$\uparrow$\\
         \hline
         GVMGen$^\dagger$&0.45&0.29&0.57&0.29&38.4&0.41&0.30&45.9&\textbf{0.91}&3.78&0.15&0.63&0.48\\
         VidMuse$^\dagger$&0.69&0.51&0.65&0.54&39.9&0.25&0.18&40.5&1.35&4.62&0.15&0.64&0.63\\
         Ours$^\ddagger$&0.66&0.43&0.63&0.43&\textbf{71.6}&0.61&0.58&69.5&0.98&\textbf{2.72}&0.17&0.64&0.48\\
         Ours&\textbf{0.87}&\textbf{0.58}&\textbf{0.89}&\textbf{0.67}&64.8&\textbf{0.92}&\textbf{0.87}&\textbf{81.9}&0.96&2.82&\textbf{0.19}&\textbf{0.66}&\textbf{0.71}\\
        \bottomrule
    \end{tabular}
\end{table*}

\begin{table}
    \caption{Subjective evaluation with 95\% confidence interval of V2M generation (top) and multi-condition controls (bottom), where $^\dagger$ indicates models trained with the two-stage strategy and multi-condition modules.}
    
    \label{sub:table}
    \begin{tabular}{cccc}
        \toprule
         Model  & OMQ$\uparrow$& MVC$\uparrow$ & UEC$\uparrow$\\
         \hline
    
         CMT&2.39$\pm$0.40&1.71$\pm$0.24&-\\
         M$^2$UGen&3.05$\pm$0.26&2.12$\pm$0.25&-\\
         Diff-BGM&2.53$\pm$0.48&1.31$\pm$0.15&-\\
         GVMGen&2.97$\pm$0.21&3.19$\pm$0.24&-\\
         VidMuse&2.72$\pm$0.17&2.84$\pm$0.25&-\\

         Ours&\textbf{3.67$\pm$0.11}&\textbf{3.67$\pm$0.21}&-\\

         \hline
         GVMGen$^\dagger$&3.05$\pm$0.24&3.43$\pm$0.22&3.19$\pm$0.16 \\
         VidMuse$^\dagger$&3.20$\pm$0.21&3.37$\pm$0.25&2.72$\pm$0.25 \\
         Ours&\textbf{3.31$\pm$0.18}&\textbf{3.56$\pm$0.25}&\textbf{3.20$\pm$0.16} \\
         
        \bottomrule
    \end{tabular}
\end{table}

\subsection{Subjective Evaluation Metrics}
For subjective evaluation, we conducted listening tests to evaluate the following aspects: Overall Music Quality (OMQ), Music-Video Correspondence (MVC) and User Expectation Conformity (UEC). OMQ measures music quality independent of the video, MVC evaluates semantic, rhythmic, and temporal consistency between music and video, and UEC assesses how well the generated music meets users’ specific expectations and preferences regarding time-varying elements. Traditional V2M generation uses OMQ and MVC, while the multi-condition guided framework additionally employs UEC.

\subsection{Comparison Models}
Owing to the lack of research specifically targeting V2M generation with multiple time-varying conditions, this paper utilizes V2M generation models as baselines, including CMT~\cite{di2021video}, M$^2$UGen~\cite{liu2023m}, Diff-BGM~\cite{li2024diff}, GVMGen~\cite{zuo2025gvmgen} and VidMuse~\cite{tian2024vidmuse}. CMT and Diff-BGM generate MIDI files using a Transformer and a diffusion model respectively, while the others generate waveform music. M$^2$UGen employs large language models to bridge music generation and visual inputs. GVMGen and VidMuse use hierarchical attentions and both local and global visual cues to generate music, respectively. For our new framework evaluation, we add our two-stage training strategy and multi-condition modules to GVMGen and VidMuse as additional baselines, allowing for fair comparison with our model.

\begin{table*}
    \caption{Ablation study of the fine-tuning stage, where "copy" represents the trainable copy layer of the PTAA's 4D-Blocks in the CGD module and "mask" stands for our masking strategy.}
    \label{obj:table4}
    \begin{tabular}{ccccccccccccc}
        \toprule
         \multirow{2}{*}{Model}&\multicolumn{2}{c}{Emotion Valence}&\multicolumn{2}{c}{Emotion Arousal}&\multicolumn{1}{c}{Melody}&\multicolumn{2}{c}{Intensity}&\multicolumn{1}{c}{Rhythm}&\multicolumn{2}{c}{Music Fidelity}&\multicolumn{2}{c}{MV-corr}\\
         \cmidrule(lr){2-3} \cmidrule(lr){4-5} \cmidrule(lr){6-6} \cmidrule(lr){7-8} \cmidrule(lr){9-9} \cmidrule(lr){10-11} \cmidrule(lr){12-13}
         \multicolumn{1}{c}{}&PCC$\uparrow$&CCC$\uparrow$&PCC$\uparrow$&CCC$\uparrow$&Acc (\%)$\uparrow$&PCC$\uparrow$&CCC$\uparrow$&F1 (\%)$\uparrow$&KLD$\downarrow$&FAD$\downarrow$&CMR$\uparrow$&TA$\uparrow$\\
         \hline
         Ours w/o. DCF&0.69&0.46&0.76&0.51&36.2&0.68&0.58&74.1&0.39&5.01&\textbf{0.71}&0.73\\
         Ours w/o. intra-block&0.82&0.57&0.88&0.66&23.2&0.13&0.09&37.7&0.19&4.47&\textbf{0.71}&0.72\\
         Ours w/o. inter-block&0.79&0.55&0.82&0.62&20.8&0.12&0.10&35.5&0.19&4.36&\textbf{0.71}&0.61\\
         \hline
         Ours w/o. CGD &0.71&0.11&0.52&0.12&44.5&\textbf{0.83}&\textbf{0.71}&21.0&0.49&8.32&0.69&0.72\\
         Ours w. copy5&0.76&0.53&0.81&0.59&19.5&0.14&0.11&32.0&0.19&4.30&\textbf{0.71}&0.71\\
         Ours w. copy3&0.84&0.55&0.89&0.66&36.2&0.16&0.12&39.0&0.19&4.42&0.70&0.72\\
         Ours w. copy2&0.79&0.53&0.83&0.63&18.8&0.09&0.07&27.9&0.19&4.38&\textbf{0.71}&0.73\\
         \hline
         Ours w/o. mask&0.73&0.16&0.61&0.16&33.8&0.56&0.48&22.1&0.53&9.66&0.68&0.72\\
         \hline
         \textbf{Ours}&\textbf{0.88}&\textbf{0.73}&\textbf{0.93}&\textbf{0.75}&\textbf{46.4}&0.71&0.55&\textbf{75.9}&\textbf{0.10}&\textbf{4.18}&\textbf{0.71}&\textbf{0.75}\\
        \bottomrule
    \end{tabular}
\end{table*}

\begin{table}
    \caption{Ablation study of the pre-training stage, where M represents masked temporal receptive field number and MR indicates reverse-sequence version of the PTAA's 4D-Blocks.}
    \label{obj:table3}
    \resizebox{0.999\linewidth}{!}{%
    \begin{tabular}{cccccc}
        \toprule
         Model  & KLD$\downarrow$& FAD$\downarrow$ & IB$\uparrow$ & CMR$\uparrow$ & TA$\uparrow$ \\
         \hline
         Ours w/o. VFA&0.28&4.75&0.12&0.71&0.72\\
         Ours w/o. VFA w. mae&0.30&\textbf{3.82}&\textbf{0.14}&0.70&0.71\\
         \hline
         Ours w/o. FGFS &0.26&4.97&0.12&0.71&0.71\\
         Ours w/o. FGFS w. clip&0.22&4.74&0.12&0.71&0.71\\
         Ours w/o. FGFS w. mae&0.22&4.81&0.13&0.71&0.71\\
         \hline
         Ours w/o. PTAA &0.28&5.06&0.11&0.72&0.72\\
         Ours w/o. PTAA w. M5 &0.23&4.74&0.12&0.71&0.71\\
         Ours w/o. PTAA w. M3 &0.25&5.18&0.12&0.71&0.72\\
         Ours w/o. PTAA w. M2 &0.26&5.32&0.11&0.73&0.72\\
         Ours w/o. PTAA w. MR &0.22&4.77&0.11&0.74&0.73\\
         \hline
         \textbf{Ours}&\textbf{0.10}&4.47&\textbf{0.14}&\textbf{0.76}&\textbf{0.75}\\
        \bottomrule
    \end{tabular}
    }
\end{table}

\subsection{Experimental Results}
This paper evaluates the performance of our model using both objective and subjective metrics, each applied separately to traditional V2M generation and multi-condition controls.
\par \textbf{Objective evaluation}. As shown in Table~\ref{obj:table1}, in traditional V2M generation tasks, our model outperforms baseline models on most objective evaluation metrics. It achieves the lowest KLD, FAD and FD scores of 0.84, 2.19 and 2.69, respectively, indicating that the music generated by our model is statistically closer to real-world music and exhibits higher perceptual quality and fidelity. Although its Density metric of 0.88 is slightly below GVMGen's 0.89, our model has demonstrated higher music fidelity in other metrics. For music richness, our model attains the highest Diversity of 68.45 and a superior Coverage of 0.80, which suggests that it can produce a wider range of musical styles. In terms of music-video correspondence, our model achieves the highest IB and TA scores, while its CMR is comparable to that of the leading baseline. This demonstrates effective semantic and temporal alignment between video and music in our model.
\par In the V2M generation with time-varying multi-condition controls, as shown in Table~\ref{obj:table2}, our method consistently outperforms two baseline models across nearly all objective metrics. It achieves enhanced emotion, melody, intensity, and rhythm control while retaining high music fidelity and music-video correspondence. We also include a variant of our model without the two-stage training strategy as an additional baseline. Although this variant shows a slight increase in melody accuracy, its performance on other metrics is noticeably lower, indicating the effectiveness of the two-stage training strategy in enhancing overall quality and controllability.

\par \textbf{Subjective evaluation}. In our user study, 20 participants, comprising 10 males and 10 females, were asked to rate 40 generated 10-second samples using a five-point Likert scale. Table~\ref{sub:table} illustrates the performance of subjective metrics for traditional V2M generation and the multi-condition guided framework. In the traditional setting, our model achieves the highest OMQ and MVC scores, reflecting its ability to generate music with superior quality and strong semantic-temporal alignment with the video. And in the multi-condition scenario, where the additional UEC metric is used, our method demonstrates top performance, indicating that it not only maintains high music quality and correspondence with the video but also greatly enhances controllability by effectively meeting user-specific expectations for time-varying elements.

\subsection{Ablation Study}
In the ablation study, we evaluated the effectiveness of each component of our model. We conducted ablation studies on both pre-training and fine-tuning stages.
\par \textbf{V2M Generation Pre-training}. Table~\ref{obj:table3} presents the performance of our model when different modules are removed or replaced. It can be observed that the overall performance of our model drops when the VFA, FGFS and PTAA modules are removed, indicating that these components are essential for the model's performance. Moreover, for the VFA module, we replaced it with pooled context-aware visual features extracted from VideoMAE V2. Although this replacement improves FAD metric, the music-video correspondence and other music fidelity metrics are still suboptimal, demonstrating the critical importance of VFA. For the FGFS module, replacing it with frame-level and context-aware visual features from CLIP and VideoMAE V2 leads to significant drops in both generative music quality and music-video correspondence. For the PTAA module, we experimented with various temporal receptive fields and a reverse-sequence version of the module to identify the most effective temporal alignment mechanism. The results indicate that the PTAA mechanism yields superior performance in music-video correspondence, both at the global and temporal levels.

\par \textbf{Multi-Condition Control Fine-tuning}. Table~\ref{obj:table4} presents the performance and controllability of our model with different modules removed or replaced while integrating multiple time-varying conditions. We can observe that the overall performance and controllability drop when the DCF, CGD modules or the masking strategy is removed, indicating that these components are essential for the model's performance and controllability. Moreover, when intra-patch or inter-patch feature fusion is not applied to the DCF module, multi-condition controllability drops significantly, which confirms that both are critical for capturing the dynamic importance of these conditions during generation. In the case of the CGD module, replacing it with other trainable copy layers of the PTAA's 4D-Blocks leads to lower overall controllability, reduced music fidelity, and diminished music-video correspondence. This suggests that incorporating a global temporal receptive field layer from the 4D-Blocks is advantageous for multi-condition fusion during music composition. It is worth noting that the model without the CGD module exhibits improved intensity controllability due to its reliance on the in-attention mechanism, but it performs poorly in other conditional controls, overall music fidelity and quality.

\section{Conclusion}
In this work, we introduce a novel multi-condition guided V2M generation framework that integrates multiple time-varying conditions with a two-stage training strategy to enhance control over music generation. In the first stage, we propose a fine-grained feature selection module and a progressive temporal alignment attention mechanism to achieve flexible temporal alignment. For the second stage, we develop a dynamic conditional fusion module and a control-guided decoder module to dynamically integrate multiple conditions and to guide music composition. Experimental results demonstrate that our method outperforms existing V2M pipelines, significantly enhancing control and aligning with user expectations.

\begin{acks}
This work is supported by the National Key Research and Development Program of China (2023YFF0904900). 
\end{acks}

\bibliographystyle{ACM-Reference-Format}
\bibliography{sample-base}

\clearpage

\appendix
\section{Details of Dataset Construction}
We use the dataset from GVMGen~\cite{zuo2025gvmgen} as our training set, which is a large-scale, high-quality collection specifically designed for video-to-music (V2M) generation. This dataset encompasses a wide range of styles, including movies, video blogs (vlogs), comics, and documentaries, with background music tailored to the video content. Additionally, it features a substantial amount of Chinese traditional music performed on over ten types of instruments, many of which cannot be adequately represented in MIDI format. The dataset is divided into training and validation sets with an 80:20 ratio.
\par For evaluation, we construct a comprehensive test set by randomly sampling and combining segments from V2M-bench~\cite{tian2024vidmuse}, SymMV~\cite{zhuo2023video}, and the GVMGen test set. V2M-bench is a benchmark dataset containing 300 video-music pairs, designed to evaluate V2M generation models across various genres, including movie trailers, advertisements, documentaries, and vlogs. SymMV comprises 1140 music videos with a total duration of 78.9 hours, a genre underrepresented in the other two datasets but similar to the MuVi-Sync dataset~\cite{kang2024video2music}. By integrating these different types of video datasets, we ensure a robust and diverse evaluation framework.
\par Moreover, to mitigate the negative impact of irrelevant human speech or singing voices in videos on V2M generation, we employ a music source separation tool~\cite{rouard2023hybrid} to process the vocals in all the music tracks mentioned above. This approach allows us to isolate and remove speech components, preserving the instrumental and background music elements that are essential for accurate V2M generation.

\section{More Implementation Details}
For three distinct video features, the dimension of patch-level fine-grained image features is 1024, while those of frame-level and context-aware visual features are 768 and 1024, respectively. In the video feature aggregation module, the feature dimension is 768, and the convolution layer has a kernel size of 3. In the fine-grained feature selection module, we apply a single 2D convolutional layer with a kernel size of 2 × 2 and a stride of 2 to downsample the patch-level fine-grained image features from 1024 to 768 dimensions. We employ an 8-head attention mechanism with 4 layers for feature selection. In the progressive temporal alignment attention module, we use 48 transformer layers with a feature dimension of 1536 as the backbone, which correspond to 12 four-layer 4D-Blocks. For audio encoding and decoding, we adopt Encodec~\cite{defossez2022high} as the default compression model for 32 kHz monophonic audio, featuring 4 codebooks of 2048 tokens. In the dynamic conditional fusion module, the patch size is set to 50. Moreover, we use top-k sampling, retaining the top 250 tokens and a temperature of 1.0 during the two training stages. 

\section{Details of Evaluation Metrics}
Due to space limitations in our main paper, we present in the following sections the details of evaluation metrics that could not be elaborated upon in the main paper.

\par \textbf{Kullback Leibler Divergence (KLD)} is a reference-dependent measure that quantifies the difference between generated and reference audio distributions. It leverages a pretrained classifier to derive class probabilities for both distributions and then computes their KL divergence. A low KLD score could indicate that the generated music has similar acoustic characteristics as the reference music, according to the classifier~\cite{agostinelli2023musiclm}.

\par \textbf{Fréchet Audio Distance (FAD)} measures the Fréchet distance between the embedding distributions of a reference audio set and the generated audio set~\cite{gui2024adapting}, for assessing audio quality. This metric evaluates how closely the generated audio resembles real audio in terms of both quality and diversity. The FAD audio encoder used in our evaluation is the VGGish model~\cite{hershey2017cnn}, which was trained on the YouTube-8M audio event dataset~\cite{abu2016youtube} for audio classification. Lower FAD values indicate higher audio plausibility~\cite{agostinelli2023musiclm}.

\par \textbf{Fréchet Distance (FD)} is a metric used to assess the similarity between generated and target samples in audio generation fields, similar to FAD. The difference from FAD is that FD employs the PANNs~\cite{kong2020panns} feature extractor, which is pretrained on the audio understanding dataset AudioSet~\cite{gemmeke2017audio}.

\par \textbf{Diversity}~\cite{li2024diff} is a metric used to evaluate the diversity of generated music. It calculates the average Euclidean distance between the music features of corresponding samples from two equally sized subsets of generated music. 
\par \textbf{Coverage}~\cite{naeem2020reliable} assesses the proportion of real samples whose neighborhoods include at least one generated sample, which reflects the richness of the generated music.

\par \textbf{ImageBind Score (IB)}~\cite{girdhar2023imagebind} evaluates how well the generated audio corresponds with the videos. Although ImageBind extends CLIP model to six modalities, only the audio and vision branches are used here. It is worth noting that ImageBind is not specifically trained on music data, which may affect the assessment of video and music consistency. However, it remains the most suitable option available for this task at present~\cite{tian2024vidmuse}. Therefore, we introduce the following two additional metrics.

\par \textbf{Cross-Modal Relevance (CMR) and Temporal Alignment (TA)}~\cite{zuo2025gvmgen} evaluate the music-video correspondence both in global and temporal aspect. The TA employs MSELoss to maximize diagonal attention, while InfoNCE Loss is used for cross-modal relevance, similar to the VMCP metric~\cite{zhuo2023video}. Higher score values indicate the music is more related and well-aligned.

\par \textbf{Pearson Correlation Coefficient (PCC) and Concordance Correlation Coefficient (CCC)} are used to evaluate the intensity and emotion controllability, similar to those in~\cite{zhang2025personalized, wu2024music}. The PCC is used to evaluate the linear correlation between the predicted values and actual values. Larger PCC values indicate a stronger positive relationship. The CCC integrates both precision and consistency, providing an enhanced measure compared to the PCC. It assesses not only the linear association but also the agreement between the means and variances of the predicted and observed values. A higher CCC value signifies that the model exhibits better performance and controllability.

\par \textbf{Rhythm F1} is used to evaluate rhythm controllability following the standard methodology~\cite{davies2009evaluation, raffel2014mir_eval} for beat/downbeat detection. It quantifies the alignment between beat and downbeat timestamps derived from the input rhythm control and those from the generated output. Timestamps are determined by applying a Hidden Markov Model (HMM) post-filter~\cite{krebs2015efficient} to the frame-wise beat and downbeat probabilities, which constitute the rhythm control signal. Finally, input and generated beat and downbeat timestamps are considered aligned if they differ by less than 70 milliseconds, as in~\cite{raffel2014mir_eval}.

\section{More Experimental Results}
\subsection{Extra Objection Evaluation}
Following~\cite{zhang2025personalized}, we also use Root Mean Square Error (RMSE) to evaluate emotion controllability in the multi-condition control framework. The RMSE metric measures the deviation between the predicted values and the actual values. A smaller RMSE value indicates higher prediction accuracy and lower model error. As shown in Table~\ref{obj1}, our model also achieves the lowest RMSE value, outperforming other models. This is consistent with the conclusion presented in the main paper.

\begin{table}[h]
    \caption{Extra objective evaluation of the V2M generation with time-varying multi-condition controls, where $^\dagger$ indicates models trained with the two-stage strategy and multi-condition modules, and $^\ddagger$ denotes the variant without the two-stage training.}
    \label{obj1}
    \begin{tabular}{ccc}
        \toprule
         \multirow{2}{*}{Model}&\multicolumn{1}{c}{Valence}&\multicolumn{1}{c}{Arousal}\\
         \cmidrule(lr){2-2} \cmidrule(lr){3-3}
         \multicolumn{1}{c}{}&RMSE$\downarrow$&RMSE$\downarrow$\\
         \hline
         GVMGen$^\dagger$&0.12&0.18\\
         VidMuse$^\dagger$&0.08&0.10\\
         Ours$^\ddagger$&0.08&0.15\\
         Ours&\textbf{0.05}&\textbf{0.06}\\
        \bottomrule
    \end{tabular}
\end{table}

\subsection{Training and Inference Time}
To provide a more comprehensive comparison, we report training and inference times in Table~\ref{obj:time1} and Table~\ref{obj:time2}, complementing Tables 1 and 2 in the main paper. As mentioned in the main paper, our model includes both a pretraining stage and a finetuning stage. However, the pretraining stage alone is sufficient for standard V2M generation. The finetuning stage, which integrates multi-condition control, freezes the pretrained parameters. Therefore, for a fair comparison with other baselines, the training time reported in Table~\ref{obj:time1} refers only to the pretraining phase.

\begin{table}[h]
    \caption{Training and inference time of V2M generation.}
    \label{obj:time1}
    \begin{tabular}{ccc}
        \toprule
         Model&Training Time/h$\downarrow$&Inference Time/s$\downarrow$\\
         \hline
         CMT&\textbf{$\approx$27}&54.75\\
         M$^2$UGen&$\approx$68&45.01\\
         Diff-BGM&$\approx$37&71.24\\
         GVMGen&$\approx$66&42.03\\
         VidMuse&$\approx$57&41.19\\
         Ours&$\approx$49&\textbf{39.49}\\
        \bottomrule
    \end{tabular}
\end{table}

\begin{table}[h]
    \caption{Training and inference time of V2M generation with time-varying multi-condition controls, where $^\dagger$ indicates models trained with the two-stage strategy and multi-condition modules, and $^\ddagger$ denotes the variant without the two-stage training strategy.}
    \label{obj:time2}
    \begin{tabular}{ccc}
         \toprule
         Model&Training Time/h$\downarrow$&Inference Time/s$\downarrow$\\
         \hline
         GVMGen$^\dagger$&$\approx$76 (66+10)&44.00\\
         VidMuse$^\dagger$&$\approx$70 (57+13)&\textbf{39.64}\\
         Ours$^\ddagger$&\textbf{$\approx$40}&41.74\\
         Ours&$\approx$61 (49+12)&40.24\\
        \bottomrule
    \end{tabular}
\end{table}

\begin{table*}[h]
    \caption{Objective evaluation of V2M generation with time-varying multi-condition created-controls, where $^\dagger$ indicates models trained with the two-stage strategy and multi-condition modules, and $^\ddagger$ denotes the variant without the two-stage training strategy.}
    \label{obj:created_table}
    \begin{tabular}{cccccccccccc}
        \toprule
         \multirow{2}{*}{Model}&\multicolumn{2}{c}{Emotion Valence}&\multicolumn{2}{c}{Emotion Arousal}&\multicolumn{1}{c}{Melody}&\multicolumn{2}{c}{Intensity}&\multicolumn{1}{c}{Rhythm}&\multicolumn{3}{c}{MV-corr}\\
         \cmidrule(lr){2-3} \cmidrule(lr){4-5} \cmidrule(lr){6-6} \cmidrule(lr){7-8} \cmidrule(lr){9-9} \cmidrule(lr){10-12}
         \multicolumn{1}{c}{}&PCC$\uparrow$&CCC$\uparrow$&PCC$\uparrow$&CCC$\uparrow$&Acc (\%)$\uparrow$&PCC$\uparrow$&CCC$\uparrow$&F1 (\%)$\uparrow$&IB$\uparrow$&CMR$\uparrow$&TA$\uparrow$\\
         \hline
         GVMGen$^\dagger$&0.19&0.02&0.37&0.04&15.5&0.26&0.06&34.0&0.14&0.64&0.54\\
         VidMuse$^\dagger$&0.16&0.01&0.27&0.04&22.2&0.66&0.12&32.6&0.14&0.65&0.55\\
         Ours$^\ddagger$&0.10&0.02&0.18&0.01&36.1&\textbf{0.75}&\textbf{0.23}&\textbf{45.4}&0.11&0.63&0.52\\
         Ours&\textbf{0.55}&\textbf{0.07}&\textbf{0.77}&\textbf{0.12}&\textbf{41.2}&0.38&0.08&42.2&\textbf{0.15}&\textbf{0.66}&\textbf{0.58}\\
        \bottomrule
    \end{tabular}
\end{table*}

\subsection{Created Controls}
\sloppy
\par Inspired by Music Controlnet~\cite{wu2024music}, we constructed a created-controls dataset containing example melodies, intensity annotations, rhythm presets, and emotion annotations that we envision creators would use during music co-creation via:
\begin{itemize}
    \item \textbf{Melody}: We collected 20 well-known music melodies (each 10 seconds long, not in our dataset) composed by Bach, Vivaldi, Mozart, Beethoven, Schubert, Mendelssohn, Bizet, as well as some pieces featuring traditional Chinese instruments, resulting in a set of 20 melody controls.
    \item \textbf{Intensity}: To simulate created intensities curves, we create 10-second-long dynamics curves as {Linear, Tanh, Cosine} functions, either vertically flipped or not, with scaled intensity ranges of {±6, ±9, ±12, ±15} decibels from the mean value of all training examples. This leads to 3 × 2 × 4 = 24 created dynamics controls.
    \item \textbf{Rhythm}: We create "rhythm presets" via selecting four music samples from our test set with different rhythmic strengths and feelings, extract their rhythm control signals, and time-stretch them using interpolation with factors {0.8, 0.9, 1.0, 1.1, 1.2} to create 20 rhythm controls.
    \item \textbf{Emotion}: To simulate diverse emotional evolutions, we draw 10-second valence and arousal curves using two types of functions: monotonically increasing and decreasing. Each curve is shifted to one of three value ranges: entirely positive, entirely negative, or crossing zero (e.g., from –0.8 to +0.8). By independently combining valence and arousal curves, we construct 6 × 6 = 36 emotion control signals for evaluation.
\end{itemize}
\par Each set of created controls is then cross-producted with 50 videos to form the final dataset of 1.0 K, 1.2 K, 1.0 K, and 1.8 K samples. Our created controls are distinct from extracted controls. The quantitative results are as shown in Table~\ref{obj:created_table}.

\subsection{Visualization}
\par
\begin{figure*}[ht]
    \includegraphics[width=0.99\textwidth]{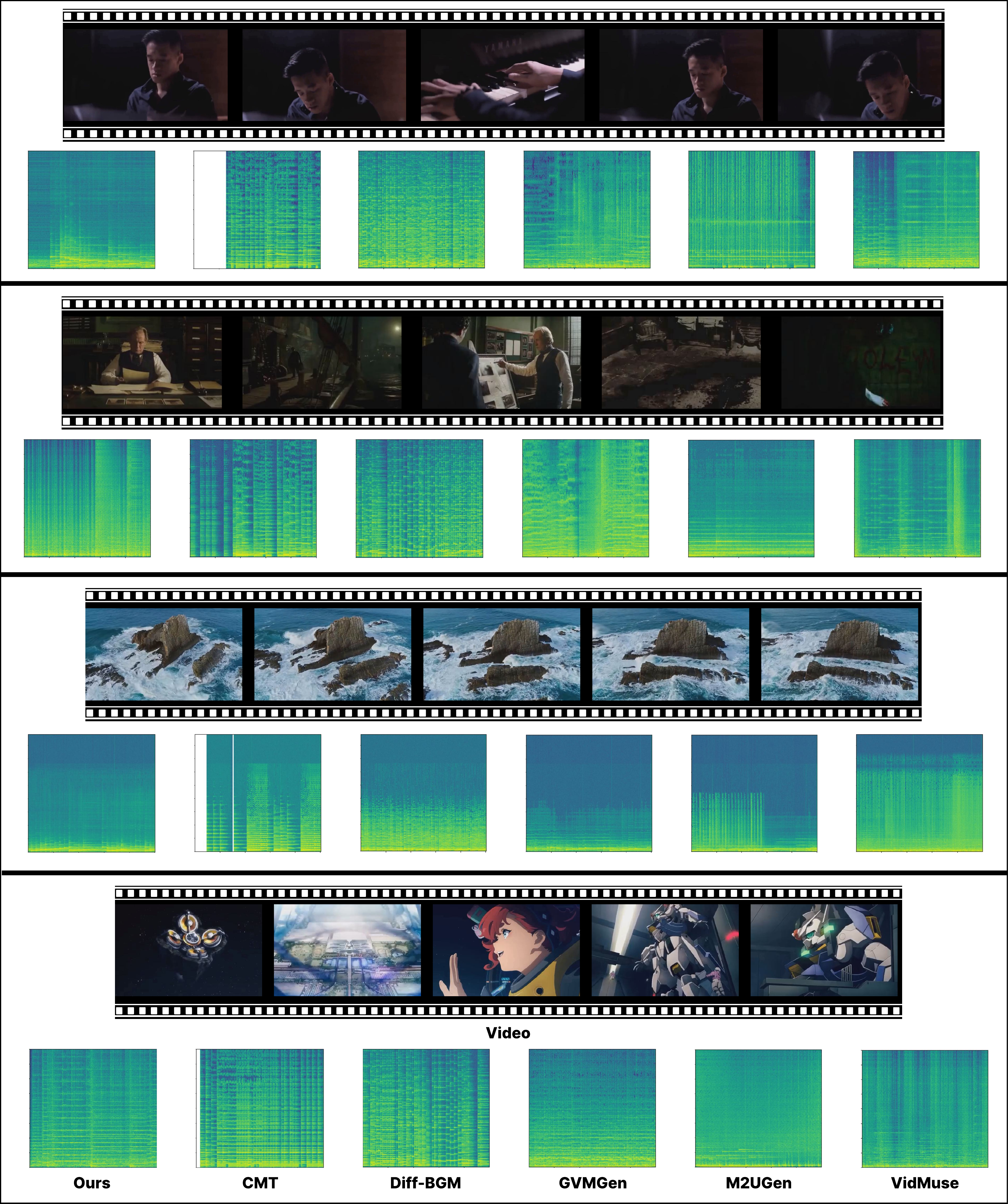}
    \caption{Spectrograms of the music generated by each model based on four types of video inputs.}
    \label{fig}
\end{figure*}

 \textbf{V2M generation framework}. As illustrated in Fig~\ref{fig}, our model produces diverse music adapted to various video types, including music videos, films, documentaries, and comics (from top to bottom). These videos exhibit distinct temporal and stylistic characteristics: the first video moves from initial stillness to a rhythmically driven piano performance, featuring a clear shift in pacing; the second maintains a consistently tense atmosphere with frequent scene transitions; the third presents a calm and steady rhythm with minimal visual changes, while the fourth exhibits a neutral tone with a visually grand and expansive style. Our model successfully aligns the generated music with both the semantic content and temporal dynamics of each video, demonstrating robustness to varying rhythms and visual types, as well as the ability to achieve flexible temporal alignment. In contrast, other models tend to produce monotonous or stylistically inconsistent music, lacking adaptability to diverse cinematic patterns.

\par
\begin{figure*}[ht]
    \includegraphics[width=0.95\textwidth]{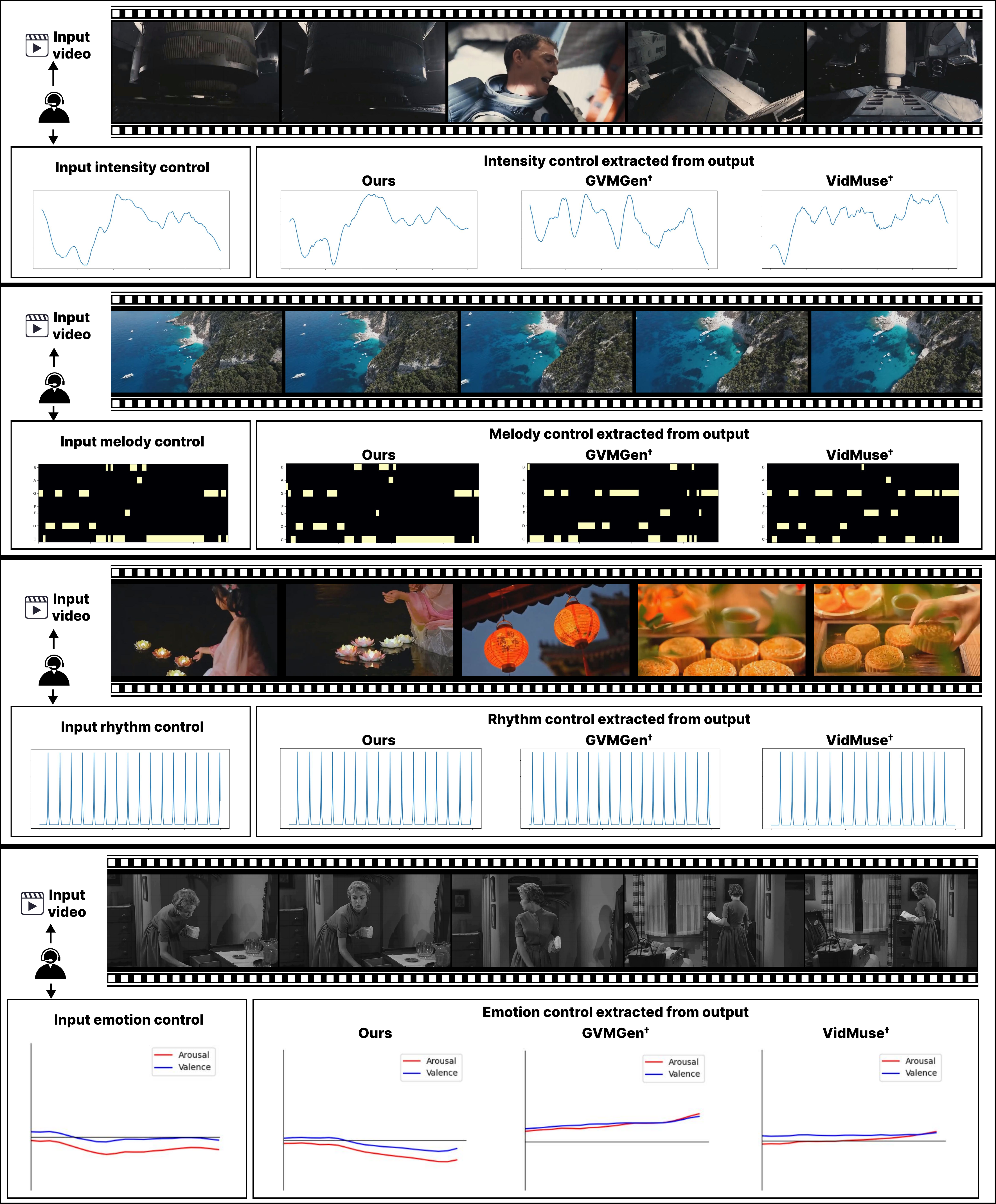}
    \caption{Examples of music generated by each model, given single time-varying control and input video. $^\dagger$ denotes models trained with our two-stage strategy and multi-condition modules.}
    \label{fig1}
\end{figure*}

 \textbf{V2M generation framework with multiple time-varying conditions}. Fig~\ref{fig1} illustrates the outputs generated by our model and two baseline models, conditioned on each of the proposed time-varying controls (i.e., intensity, melody, rhythm, or emotion) and input videos. Our model effectively attends to the varying control signals, ensuring that the generated outputs consistently reflect these controls while aligning with the video's semantic content and temporal dynamics. This demonstrates both the high controllability of our framework and its robust capability in V2M generation. In contrast, even with the incorporation of our strategy and module, other models still exhibit limitations in their controllability, with generated outputs lacking alignment with the specific conditions or visual inputs.

\par
\begin{figure*}[ht]
    \includegraphics[width=\textwidth]{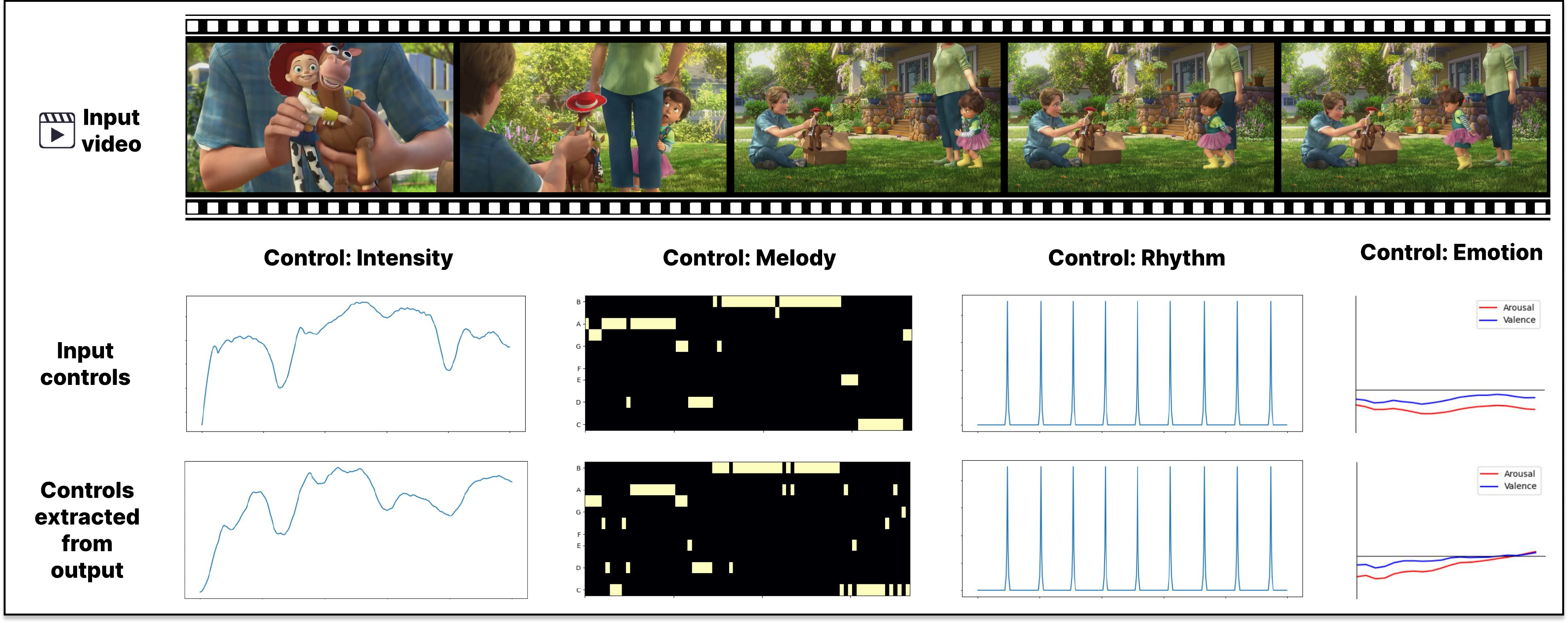}
    \caption{An example of music generation guided by input video and multiple user-specified time-varying conditions.}
    \label{fig2}
\end{figure*}

\begin{figure*}[ht]
    \includegraphics[width=\textwidth]{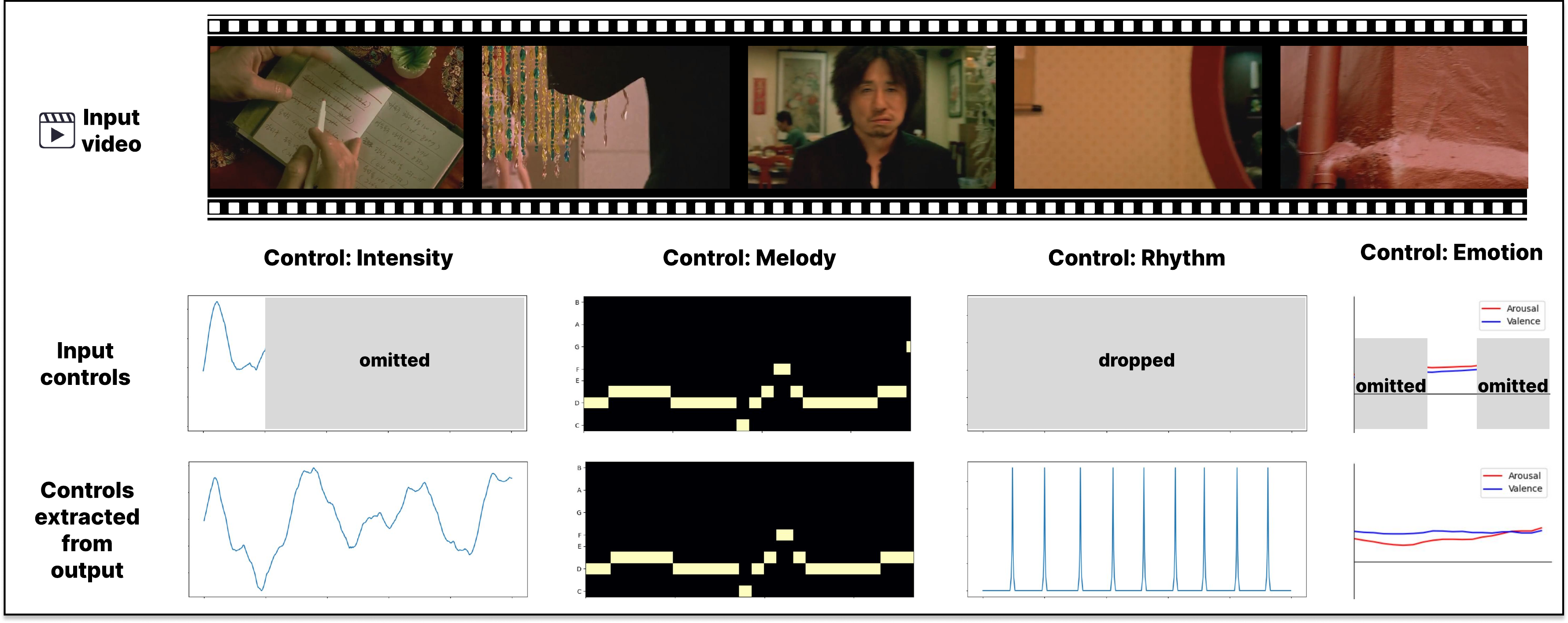}
    \caption{An example of music generation guided by video input and multiple time-varying conditions that are partially specified by users in terms of both content and temporal dynamics.}
    \label{fig3}
\end{figure*}

 Moreover, as detailed in our main paper, our model can generate music based on various combinations of four control conditions, as well as handle scenarios where parts of an individual condition's temporal sequence are masked. Fig~\ref{fig2} illustrates a case where all time-varying control signals are fully specified. The generated music effectively integrates multiple time-varying conditions and video information, closely aligning with the composite guidance. The control attributes extracted from the output exhibit high temporal consistency with the input controls, and the resulting music remains well synchronized with the video content in both its dynamics and semantic content. Fig~\ref{fig3} shows a generation example with partially specified controls, where some conditions or their temporal segments are omitted. Our model fills in the missing parts with musically appropriate content, maintaining a consistent style and musical creativity. This enables users to guide the generation process flexibly, without the need to provide complete control sequences.

\section{Limitation and Discussion}
In this work, we introduced multiple time-varying musical and emotional controls for the first time in the V2M generation task. While this novel framework significantly enhances controllability over the generated music, it also presents some limitations. Specifically, when multiple conditions are simultaneously provided, it can be challenging for the model to fully comply with all control signals. Additionally, conflicts between different control conditions and the video style can adversely affect the music generation, leading to suboptimal outputs. In the future, we will refine the model's ability to handle such conflicting conditions more effectively to better align with user preferences.

\end{document}